\documentclass[conference]{IEEEtran}
\IEEEoverridecommandlockouts

\usepackage[switch]{lineno}

\usepackage{amsmath,amsfonts}
\usepackage{algorithmic}
\usepackage{algorithm}
\usepackage{array}
\usepackage[caption=false,font=normalsize,labelfont=sf,textfont=sf]{subfig}
\usepackage{textcomp}
\usepackage{stfloats}
\usepackage{url}
\usepackage{verbatim}
\usepackage{graphicx}
\usepackage{cite}
\usepackage{hyperref}
\usepackage{multirow}
\usepackage{amsmath}
\usepackage{forest}
\usepackage{listings,jvlisting}
\usepackage{makecell}

\def\cm{\checkmark}
\def\BibTeX{{\rm B\kern-.05em{\sc i\kern-.025em b}\kern-.08em
    T\kern-.1667em\lower.7ex\hbox{E}\kern-.125emX}}

\begin{document}

\title{BA-TRACE: Boundary-Aware Trace Reconstruction for Scenario-Based Evaluation of Mixed AUTOSAR Adaptive and ROS 2 Vehicular Embedded Systems}






\author{
\IEEEauthorblockN{
Shunsuke Ito\IEEEauthorrefmark{1},
Ryudai Iwakami\IEEEauthorrefmark{1},
Hiroyuki Hanyu\IEEEauthorrefmark{2},
Tasuku Ishigooka\IEEEauthorrefmark{2},
and Takuya Azumi\IEEEauthorrefmark{3}
}
\IEEEauthorblockA{
\IEEEauthorrefmark{1}
\textit{Graduate School of Science and Engineering, Saitama University}
}
\IEEEauthorblockA{
\IEEEauthorrefmark{2}
\textit{Technology Development Functional Division, Astemo, Ltd.}
}
\IEEEauthorblockA{
\IEEEauthorrefmark{3}
\textit{Academic Association (Graduate School of Science and Engineering), Saitama University}
}
}


\maketitle

\begin{abstract}
Modern vehicular embedded systems increasingly combine ROS~2-based autonomous-driving stacks with AUTOSAR Adaptive Platform (AUTOSAR~AP). Such mixed stacks make scenario-based evaluation hard to interpret because execution paths cross DDS--SOME/IP middleware boundaries between ROS~2 and AUTOSAR~AP. Existing simulators and tracing tools execute scenarios or collect platform-local traces but cannot reconstruct cross-domain data flows.
This paper presents \emph{BA-TRACE}, a boundary-aware trace reconstruction framework for scenario-based evaluation of mixed AUTOSAR~AP and ROS~2 vehicular embedded systems. BA-TRACE combines ROS~2 trace events, AUTOSAR \texttt{ara::log} events, ARXML-derived structural dependencies, and bridge-level instrumentation to reconstruct an end-to-end execution graph across the DDS--SOME/IP boundary. A case study with an AWSIM/OpenSCENARIO-based object-detection and braking scenario shows that BA-TRACE reconstructs the expected cross-platform path and exposes boundary-specific latency such as point-cloud transfer overhead. The reconstructed topology is used as evidence of traceability, not as proof of behavioral correctness or safety.
\end{abstract}


\section{Introduction}
\label{sec:introduction}

Autonomous driving technologies are moving from laboratory prototypes toward practical deployment, and credible safety validation is becoming increasingly important.
Developers must evaluate not only nominal driving behavior but also rare and hazardous edge cases (e.g., cut-in with sudden stop) involving complex interactions with surrounding traffic participants~\cite{Riedmaier2020survey}.
Scenario-based validation has emerged as a promising approach to this challenge~\cite{ding2022survey}, and standard scenario description languages such as OpenSCENARIO~\cite{OpenSCENARIO} enable reusable and structured test design.

Software-Defined Vehicles (SDVs) are increasing the heterogeneity of automotive software environments.
In practice, development and validation are increasingly conducted in mixed environments combining AUTOSAR Adaptive Platform (AUTOSAR~AP), a production-grade automotive foundation, with Robot Operating System 2 (ROS~2)~\cite{comparison}.
Such mixed software stacks make scenario-based evaluation difficult to interpret because execution paths cross heterogeneous middleware boundaries, namely DDS-based ROS~2 communication and SOME/IP-based AUTOSAR~AP communication.
Existing simulation and tracing tools can execute driving scenarios or collect platform-local traces, but do not reconstruct how scenario-triggered data flows across these execution domains.

This paper does not propose a new simulator or scenario description language.
Instead, it proposes \emph{BA-TRACE}, a boundary-aware trace reconstruction workflow making scenario-based evaluation diagnostically interpretable for mixed AUTOSAR~AP and ROS~2 vehicular embedded systems.
BA-TRACE combines ROS~2 trace events, AUTOSAR \texttt{ara::log} events, ARXML-derived structural dependencies, and bridge-level instrumentation to reconstruct an end-to-end execution graph across the DDS--SOME/IP boundary.

Based on this objective, the following research questions (RQs) are formulated:
\begin{itemize}
    \item \textbf{RQ1:} What traceability gap arises when scenario-based evaluation is applied to mixed AUTOSAR~AP and ROS~2 vehicular embedded systems?
    \item \textbf{RQ2:} Can boundary-aware trace reconstruction recover the expected end-to-end execution topology across DDS and SOME/IP communication domains?
    \item \textbf{RQ3:} How does the reconstructed cross-platform execution graph support timing interpretation of scenario-triggered behavior?
\end{itemize}

The contributions of this paper are summarized as follows:
\begin{itemize}
    \item We formulate the traceability problem in scenario-based evaluation of mixed AUTOSAR~AP--ROS~2 systems, identifying semantic and timing gaps across DDS, SOME/IP, and bridge-level boundaries.

    \item We propose \emph{BA-TRACE}, a boundary-aware trace reconstruction workflow combining ROS~2 traces, AUTOSAR \texttt{ara::log} events, ARXML-derived dependencies, and bridge-level instrumentation to reconstruct an end-to-end execution graph across the DDS--SOME/IP boundary.

    \item We implement BA-TRACE in a vehicular embedded testbed using Autoware, AUTOSAR~AP, AWSIM, OpenSCENARIO, and an instrumented bridge converter. An edge-case object-detection and braking scenario shows the reconstructed graph exposes cross-platform paths and boundary-specific latency.
\end{itemize}

The reconstructed topology is used as evidence of traceability, not as proof of behavioral correctness or safety.

The remainder of this paper is organized as follows.
Section~\ref{chap:system_model} describes the technical background of AUTOSAR~AP and ROS~2.
Section~\ref{chap:design_and_implementation} presents the proposed framework, and Section~\ref{chap:evaluation} reports the evaluation results.
Section~\ref{chap:lessons_learned} discusses lessons learned, and Section~\ref{chap:related_work} reviews related work.
Finally, a brief conclusion is given.

\section{System Model}
\label{chap:system_model}

\begin{figure}[t]
\centering
    \includegraphics[width=0.9\linewidth]{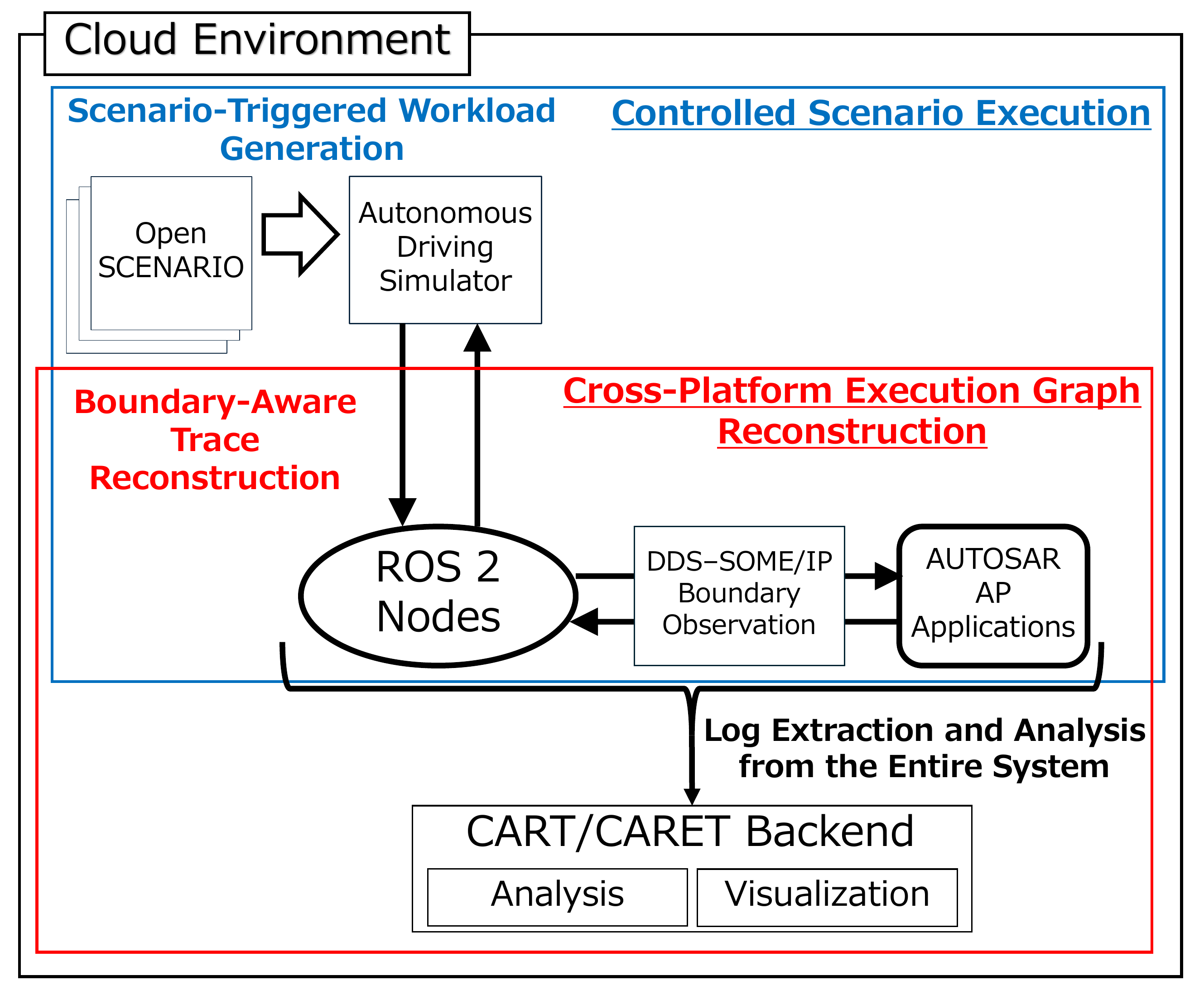}
    \caption{Architecture of the BA-TRACE framework.}
    \label{fig:system_model}
    \vspace{-4mm}
\end{figure}

The BA-TRACE framework targets mixed AUTOSAR~AP--ROS~2 vehicular embedded systems, where scenario-triggered execution paths cross the DDS--SOME/IP boundary at a bridge converter. Its overall architecture is shown in Fig.~\ref{fig:system_model}. 
The framework integrates five technical building blocks: OpenSCENARIO as the scenario specification language, AWSIM as the autonomous driving simulator, ROS~2-based Autoware and AUTOSAR~AP applications as the heterogeneous mixed stack, and the CART~\cite{CART}/CARET~\cite{CARET} pipeline as the trace correlation and visualization backend used to render the reconstructed cross-platform execution graph. The remainder of this section reviews each building block, starting with OpenSCENARIO as the entry point of the workflow.

\subsection{OpenSCENARIO}
\label{sec:OpenSCENARIO}

In scenario-based evaluation of autonomous driving systems, OpenSCENARIO is used as a standard description format for defining test cases~\cite{OpenSCENARIO}. Critical traffic situations and edge cases, such as sudden lane changes or pedestrians darting out, are described concretely as XML-based scenarios. As an open standard independent of specific tools, scenarios written once can be reused across various simulators, enabling efficient and reproducible testing.

OpenSCENARIO uses a hierarchical structure of Story, Act, Maneuver, and Event, and supports precise behavior definition via tags such as \texttt{<LaneChangeAction>} (with attributes specifying, for example, completion within 0.5 seconds for aggressive cut-in) and \texttt{<StartTrigger>} (firing at a deterministic simulation time), enabling reproducible and platform-independent test cases.

\subsection{Autonomous Driving Simulator}
\label{sec:ADS}
Autonomous driving simulators provide a virtual environment with road structures, traffic participants, and weather/lighting conditions, and are equipped with sensor models such as LiDAR, cameras, radar, and GNSS.
They model the ego-vehicle and surrounding entities and reproduce diverse driving scenarios in a controlled manner, including situations that are difficult or dangerous to reproduce in the real world.
Representative examples include CARLA~\cite{CARLA}, which features high-fidelity rendering on Unreal Engine, and AWSIM~\cite{AWSIM}, which is Unity-based, lightweight, and suitable for cloud environments; trade-offs therefore exist among realism, flexibility, and computational cost.
AWSIM is the simulator used in BA-TRACE.

Most autonomous driving simulators are independent systems and do not natively use the DDS-based communication of ROS~2; they expose simulator-specific client--server APIs in Python or C++.
A dedicated bridge is therefore required to integrate them with ROS~2-based stacks.
The bridge converts simulator-internal data into ROS~2 messages and synchronizes control commands back to the simulator, allowing the simulator to function as a virtual sensor input source for the autonomous driving stack.

\begin{figure}[t]
\centering
    \includegraphics[width=0.8\linewidth]{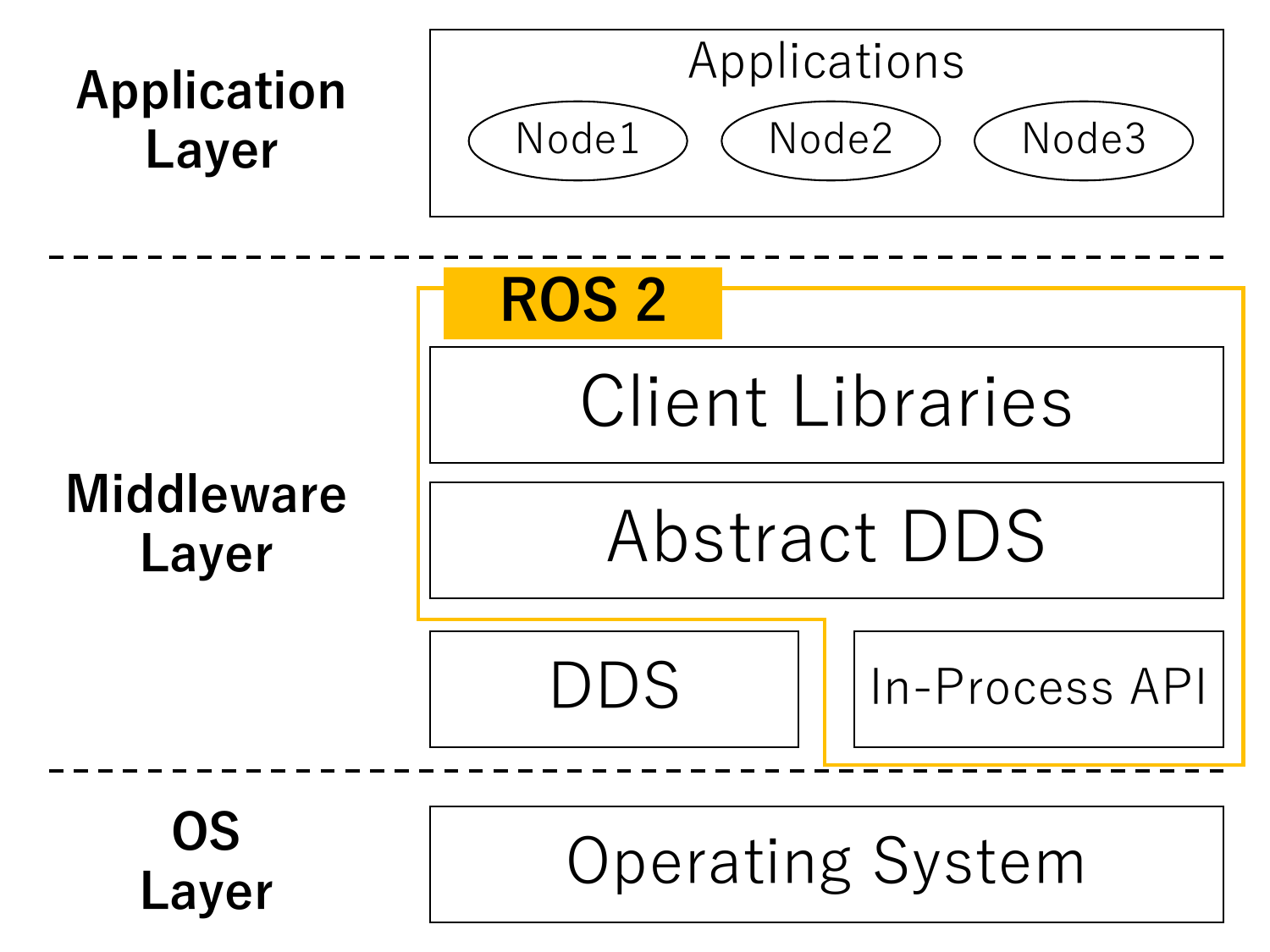}
    \caption{ROS~2 architecture.}
    \label{fig:ROS_2_simple_architecture}
    \vspace{-4mm}
\end{figure}

\subsection{Robot Operating System 2 (ROS~2)}
\label{sec:ROS2}



ROS~2 is open-source middleware widely used in autonomous driving software and industrial systems~\cite{comparison} and meets real-time constraints through DDS. The three-layer ROS~2 architecture (Fig.~\ref{fig:ROS_2_simple_architecture}) comprises application, middleware, and OS layers; the application hosts user code and ROS~2 nodes, while the middleware uses DDS for inter-node communication. ROS~2 \emph{nodes} exchange data through \emph{topics} via a publish/subscribe model. This loose coupling enables efficient data sharing and a broad tool ecosystem, including the tracing tools ros2\_tracing~\cite{ros2_tracing} and CARET, both used by BA-TRACE on the ROS~2 side.

\subsection{AUTOSAR Adaptive Platform (AUTOSAR~AP)}
\label{sec:AUTOSARAP}

\begin{figure}[t]
\centering
    \includegraphics[width=0.9\linewidth]{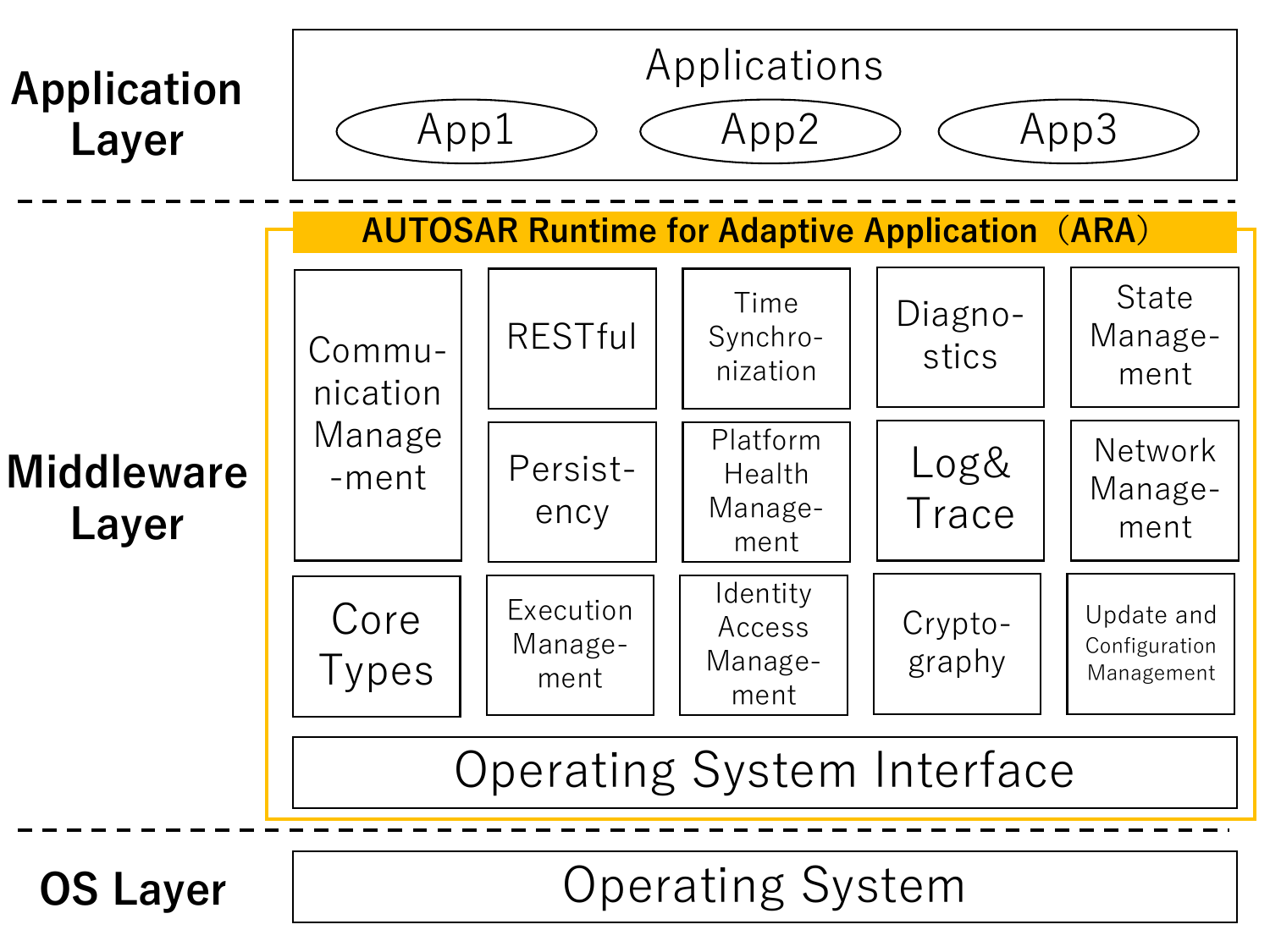}
    \caption{AUTOSAR~AP architecture.}
    \label{fig:AUTOSAR_AP_simple_architecture}
    \vspace{-4mm}
\end{figure}

AUTOSAR~AP is a software architecture for next-generation automotive embedded systems~\cite{comparison}.
Compared to ROS~2 with DDS, AUTOSAR~AP uses SOME/IP (Scalable service-Oriented MiddlewarE over IP) as its main communication protocol.
Unlike ROS~2, AUTOSAR~AP requires a license, is less openly documented, and can be more difficult to handle due to extensive standardization.
Its architecture (Fig.~\ref{fig:AUTOSAR_AP_simple_architecture}) is layered into application, middleware, and OS, similar to ROS~2~\cite{comparison}.

The middleware, AUTOSAR Runtime for Adaptive Application (ARA), provides core services including the standard logging interface \texttt{ara::log}.
Adaptive Applications (AAs) communicate via service requests under a Service-Oriented Architecture (SOA) supported by SOME/IP.
The data types and interfaces handled by AAs, the SOME/IP configurations, and the inter-AA structural dependencies are all defined in AUTOSAR XML (ARXML), which BA-TRACE consumes as its static structural source for trace reconstruction.

\begin{figure}[t]
\centering
    \includegraphics[width=0.9\linewidth]{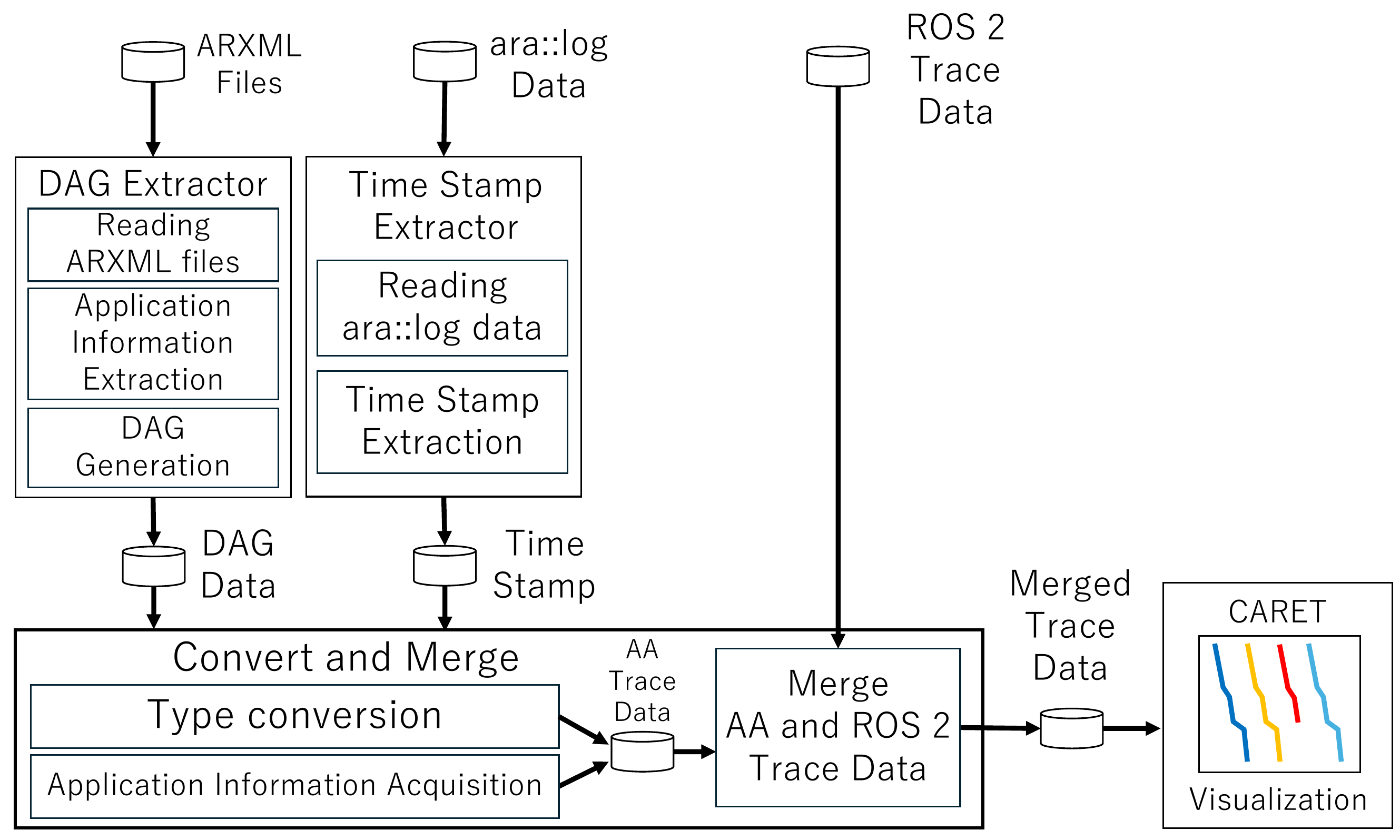}
    \caption{CART structure within BA-TRACE.}
    \label{fig:CART_structure}
    \vspace{-4mm}
\end{figure}

\subsection{CART (Combined AUTOSAR~AP and ROS~2 Tracing Framework)}
\label{sec:aralog}
CART (Combined AUTOSAR~AP and ROS~2 Tracing)~\cite{CART} is the trace correlation backend used by BA-TRACE for mixed AUTOSAR~AP and ROS~2 systems. CART first extracts structural dependencies among Adaptive Applications from ARXML files and represents them as a directed acyclic graph (DAG). It then reads execution timestamps from \texttt{ara::log} and converts AUTOSAR-side events into Common Trace Format (CTF)-based trace data compatible with ROS~2 analysis tools. These traces are merged with native ROS~2 trace data collected via ros2\_tracing, allowing the entire cross-platform execution flow to be analyzed and visualized in a unified manner. Within BA-TRACE, CART is the component that fuses ARXML-derived static structure with dynamic ROS~2 and AUTOSAR-side events, producing the merged trace from which the reconstructed cross-platform execution graph and latency breakdown are derived.
The open-source implementation of CART is publicly available on GitHub~\cite{CART_github}.

\subsection{CARET}
\label{sec:caret}



CARET (Chain-Aware ROS~2 Evaluation Tool)~\cite{CARET} is a performance analysis tool for ROS~2 applications and serves as the BA-TRACE frontend for visualizing and analyzing latency in the merged trace produced by CART (Fig.~\ref{fig:CART_structure}). 
By adding tracepoints to ROS~2 tracing, CARET measures inter-node communication and callback execution times and message-flow dependencies without modifying application code, enabling bottleneck and lost-message detection. 
CARET presents latency-distribution histograms, tables of minimum, average, and maximum latency, and message-flow graphs, which BA-TRACE reuses to render the reconstructed cross-platform execution graph and the latency breakdown reported in Section~\ref{chap:evaluation}.
\section{Design and Implementation}
\label{chap:design_and_implementation}

\subsection{Overview of BA-TRACE}

The BA-TRACE framework, whose architecture is shown in Fig.~\ref{fig:system_model}, integrates a scenario-driven simulation testbed with a boundary-aware trace reconstruction workflow.
The testbed executes reproducible OpenSCENARIO-defined scenarios in a cloud-native environment, and the workflow combines ROS~2 trace events, AUTOSAR \texttt{ara::log} events, ARXML-derived structural dependencies, and bridge-level instrumentation to reconstruct an end-to-end execution graph across the DDS--SOME/IP boundary.
The remainder of this section first describes the trace reconstruction workflow (Section~\ref{sec:reconstruction_workflow}) and the bridge-level instrumentation that anchors it (Section~\ref{sec:bridge_instrumentation}), and then the cloud simulation testbed (Section~\ref{sec:Validation_Process}) and the mixed Autoware--AUTOSAR~AP implementation (Section~\ref{sec:mixed_implementation}) on which BA-TRACE is exercised.

\subsection{Boundary-Aware Trace Reconstruction Workflow}
\label{sec:reconstruction_workflow}

BA-TRACE reconstructs an end-to-end execution graph by correlating four sources: ROS~2 trace events, AUTOSAR \texttt{ara::log} events, ARXML-derived structural dependencies, and bridge-level boundary events.
The Bridge Converter is treated as an explicit observation point because it is the only component that simultaneously observes ROS~2-side DDS topic semantics and AUTOSAR-side SOME/IP service semantics.

Concretely, BA-TRACE reconstructs the execution graph in four steps:
(1) extracting ROS~2 callback and communication events from ros2\_tracing;
(2) extracting AUTOSAR-side structural dependencies from ARXML and runtime timestamps from \texttt{ara::log};
(3) correlating DDS-side and SOME/IP-side events using bridge-level instrumentation and correlation identifiers; and
(4) merging the ROS~2, AUTOSAR~AP, and boundary edges into a single cross-platform execution graph from which the latency breakdown is derived.

\subsection{Bridge-Level Instrumentation at the DDS--SOME/IP \\ Boundary}
\label{sec:bridge_instrumentation}

A Bridge Converter (BC) handles bidirectional translation between the DDS protocol used in ROS~2 and the SOME/IP protocol used in AUTOSAR~AP, including data-type serialization.
The Bridge Converter is the only component that simultaneously observes both ROS~2-side topic semantics and AUTOSAR-side SOME/IP service semantics.
BA-TRACE therefore instruments it as a boundary observation point: in addition to protocol and data conversion, the bridge captures timing information at the DDS--SOME/IP boundary and embeds correlation identifiers that link DDS-side and SOME/IP-side events.
This boundary-level instrumentation is the technical core of BA-TRACE and is what makes the cross-platform execution graph reconstructible.

\subsection{Scenario-Driven Cloud Simulation Testbed}
\label{sec:Validation_Process}

Scenario-driven evaluation of BA-TRACE relies on a cloud-native testbed combining OpenSCENARIO as the scenario specification language with AWSIM as the autonomous driving simulator.


\subsubsection{Cloud-Native Evaluation Environment}
\label{subsec:Cloud-Native Evaluation Environment}
The foundation of the testbed is a cloud-native evaluation environment, aligned with the Scalable Open Architecture for Embedded Edge (SOAFEE) framework that promotes cloud-native development for SDVs.
The primary rationale for adopting a cloud environment is reproducibility and resource elasticity, which are often compromised in on-premise setups due to ``environment drift'', subtle discrepancies in OS versions, drivers, or library dependencies that lead to the ``it works on my machine'' paradox.
Cloud infrastructure resolves this by enabling pre-configured machine images: cloning a standardized image instantiates identical instances for every test run, eliminating environmental variability and keeping evaluation results consistent regardless of when or by whom the tests are executed.
This cloning capability also synergizes with resource elasticity, allowing scenarios to be parallelized across multiple instances (scale-out) or migrated to higher-performance instance types (scale-up), which is crucial for future integration with Continuous Integration pipelines.

\subsubsection{Adoption of OpenSCENARIO for Reproducible \\ Validation}
\label{subsec:Adoption of OpenSCENARIO for Reproducible Validation}
ASAM OpenSCENARIO 1.x is adopted as the standard format for defining test cases that cover both routine driving situations and rare safety-critical edge cases.
The decision to adopt OpenSCENARIO, as opposed to simulator-specific proprietary scripts or random traffic generation, is driven by interoperability and reproducibility.
Random traffic generation lacks the determinism required to consistently reproduce specific hazardous situations such as a pedestrian darting out at a precise timing, while proprietary scripts lock validation assets to a specific simulator and hinder cross-platform verification.
By using OpenSCENARIO, BA-TRACE keeps scenario assets portable and ensures that complex interaction events can be defined with strict temporal and spatial precision.
At runtime, the simulator parses \texttt{.xosc} files to spawn entities and set initial conditions, monitors triggers such as ego-vehicle speed or position, and dynamically executes maneuvers, supporting controlled replay of logically equivalent scenario conditions and providing a scenario stimulus for subsequent trace reconstruction.

\subsection{Mixed Autoware--AUTOSAR~AP Implementation}
\label{sec:mixed_implementation}

The evaluation target is a mixed environment in which Autoware provides the ROS~2-side autonomous driving stack and AUTOSAR Adaptive Applications carry functions deployed on the AUTOSAR Adaptive Platform.
The simulator transmits sensor inputs (e.g., LiDAR point clouds) to Autoware via a dedicated simulator-to-ROS~2 bridge, and the Bridge Converter described in Section~\ref{sec:bridge_instrumentation} handles the DDS--SOME/IP boundary between Autoware nodes and the AUTOSAR-side AAs.
The specific deployment used for evaluation, in which object detection is hosted by an AUTOSAR Adaptive Application, is detailed in Section~\ref{chap:evaluation}.

\section{Evaluation}
\label{chap:evaluation}

The BA-TRACE framework is implemented as described below and evaluated from three perspectives aligned with the research questions in Section~\ref{sec:introduction}.
First, the evaluation examines the traceability gap that arises when a scenario-triggered execution path crosses the DDS--SOME/IP boundary in a mixed AUTOSAR~AP--ROS~2 stack (RQ1).
Second, it examines whether boundary-aware trace reconstruction recovers the expected end-to-end execution topology across DDS and SOME/IP communication domains (RQ2).
Third, it examines whether the reconstructed execution graph supports timing interpretation of scenario-triggered behavior (RQ3).
The reported results should be interpreted as evidence of diagnostic traceability and timing interpretability, not as proof of behavioral correctness or safety.

\begin{figure}[t]
\centering
\includegraphics[width=1.0\linewidth]{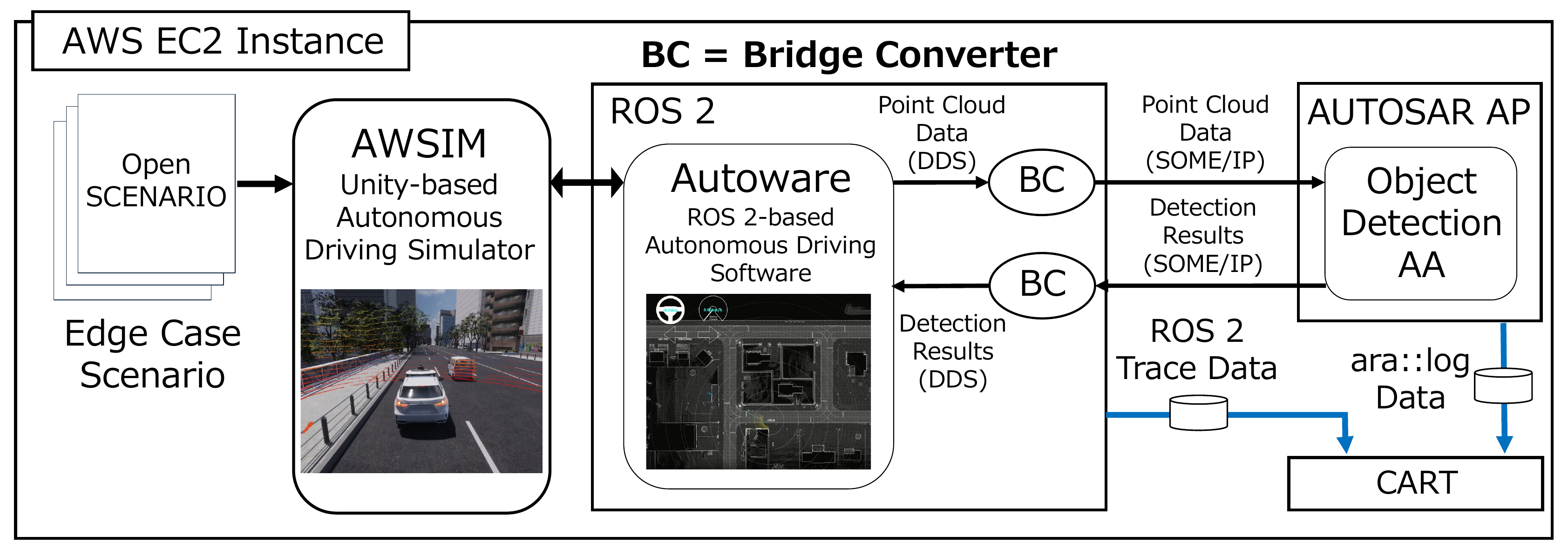}
\caption{Implementation of the proposed platform for evaluation.}
\label{fig:proposed_platform_for_evaluation}
\vspace{-4mm}
\end{figure}

\subsection{Evaluation Platform Details}

The detailed implementation of the evaluation platform is shown in Fig.~\ref{fig:proposed_platform_for_evaluation}.
AWS (Amazon Web Services)~\cite{AWS} was selected for the cloud environment, and the testbed was implemented on EC2 instances. The Bridge Converter employs an open-source implementation~\cite{BC_github,BC_AWS} enhanced with the improvements described in Section~\ref{chap:design_and_implementation}.
AWSIM was selected as the autonomous driving simulator for its native ROS~2/Autoware integration and direct OpenSCENARIO ingestion.

The simulator and the autonomous driving system were deployed on a single high-performance instance to remove network jitter and bandwidth bottlenecks inherent in distributed setups, which would otherwise introduce noise into the timing measurements.
This lets boundary-specific overhead (such as the 7.62~ms point-cloud transfer cost reported in Section~\ref{sec:4.2}) be attributed to the software stack itself rather than to external network variability, a methodological choice for diagnostic clarity, not a substitute for distributed-deployment evaluation.

Logs are collected from the executing autonomous driving system using CART (Combined AUTOSAR~AP and ROS~2 Tracing framework), the trace correlation backend used by BA-TRACE.
Because AWSIM and the autonomous driving stack require a GPU and significant memory, a high-performance instance was adopted (Table~\ref{tab:envs}).

The Bridge Converter, adopted from the open-source implementation~\cite{BC_github}, was modified to enable boundary-aware tracing across the heterogeneous DDS--SOME/IP boundary.
Since its original SOME/IP transmission and reception modules do not natively generate traceable events via \texttt{ara::log} or ros2\_tracing, tracepoints were added only at the bridge to capture protocol-boundary timing and embed correlation identifiers, without modifying the core logic of the AUTOSAR Adaptive Application or the ROS~2 nodes.

Autoware served as the base, with object detection handled by an AUTOSAR Adaptive Application (AA), creating a mixed AUTOSAR~AP and ROS~2 environment.
As illustrated in Fig.~\ref{fig:proposed_platform_for_evaluation}, point cloud data acquired from AWSIM is transmitted via DDS to the Bridge Converter, which converts the data to SOME/IP and forwards it to the Object Detection AA on the AUTOSAR~AP side.
The detection results are returned to Autoware through the Bridge Converter by converting them back from SOME/IP to DDS, and the system issues a stop command upon detecting an object in front.
To exercise this logic, an edge-case scenario was defined in OpenSCENARIO: on a two-lane main road, an NPC vehicle overtakes the ego-vehicle from the right lane, cuts in front of it, and immediately performs a sudden stop.
This setup verifies object detection responsiveness and braking maneuver execution under aggressive cut-in and sudden deceleration.
The software versions used in this evaluation are listed in Table~\ref{tab:version}.

\begin{table}[t]
\centering
\scriptsize
\begin{minipage}[t]{0.48\linewidth}
\centering
\caption{Evaluation Instance Configuration}
\label{tab:envs}
\begin{tabular}{|l|c|}
\hline
Instance Type & g5.8xlarge \\ \hline
CPU & \makecell{AMD EPYC 7R32\\32 cores} \\ \hline
RAM & 128 GB \\ \hline
GPU & \makecell{NVIDIA A10G\\Tensor Core GPU} \\ \hline
VRAM & 24 GB \\ \hline
\end{tabular}
\end{minipage}
\hfill
\begin{minipage}[t]{0.48\linewidth}
\centering
\caption{Used software versions}
\label{tab:version}
\begin{tabular}{|l|c|}
\hline
Software & Version \\ \hline
OS & Ubuntu 22.04 \\ \hline
AUTOSAR~AP & R20-11 \\ \hline
ROS~2 & Humble Hawksbill \\ \hline
Autoware & 0.45.1 \\ \hline
AWSIM & v2.0.1 \\ \hline
LTTng & 2.13 \\ \hline
ros2\_tracing & v0.3.0 \\ \hline
CARET & v0.6.1 \\ \hline
\end{tabular}
\end{minipage}
\vspace{-8mm}
\end{table}

\subsection{Boundary-Aware Trace Reconstruction in a Practical Simulation Environment}
\label{sec:4.2}

In this section, BA-TRACE is evaluated using the simulation testbed described in the previous section.
By executing the autonomous driving system under the defined edge-case scenario, three properties are examined: (i) the traceability gap that emerges when scenario-triggered data crosses the DDS--SOME/IP boundary, (ii) the recovery of the expected end-to-end execution topology by boundary-aware trace reconstruction, and (iii) the timing interpretation of scenario-triggered behavior via the reconstructed execution graph.

\subsubsection{Scenario-Triggered Execution and the Traceability Gap}

The testbed was exercised under the edge-case cut-in scenario described in the previous subsection.
AWSIM reproduced the specified traffic situation, and the mixed autonomous driving stack operated on the same cloud instance while processing sensor data and object detection results across the ROS~2--AUTOSAR~AP boundary, providing a stable testbed condition under which scenario-triggered execution can be observed.

This result establishes the testbed condition needed to investigate RQ1.
Concretely, the scenario-triggered data flow visibly crosses the DDS--SOME/IP boundary at the bridge converter, the point at which ROS~2 trace events or AUTOSAR \texttt{ara::log} events alone become insufficient to describe the cross-platform path.
The traceability gap motivated in Section~\ref{sec:introduction} therefore manifests in the testbed as a concrete observation problem, which the next subsubsection addresses by topology-level reconstruction.

\subsubsection{Topology Reconstruction and Timing Interpretation}

The collected logs are fed into BA-TRACE to reconstruct the cross-platform execution graph and measure the end-to-end latency of the mixed Autoware--AUTOSAR~AP system.
End-to-end latency is defined as the duration from the reception of point cloud data by the ROS~2 input node~(\textit{/pointcloud\_downsampler}) to the issuance of the corresponding control command by the actuation node~(\textit{/relay\_node}), which immediately forwards the command to the simulator interface.
Topology coverage is defined as the edge coverage between the expected graph (from design artifacts) and the observed graph reconstructed by CARET from merged traces.

Here, $E_{\mathrm{expected}}$ denotes the set of dependency edges in the complete end-to-end path, defined by the design artifacts~(i.e., ARXML for AUTOSAR~AP and node configurations for ROS~2) prior to execution.
Meanwhile, $E_{\mathrm{observed}}$ denotes the set of edges actually extracted from the merged runtime traces.
Accordingly, the coverage metric is defined as:
\begin{equation}
\mathrm{Coverage} =
\frac{\lvert E_{\mathrm{observed}} \cap E_{\mathrm{expected}} \rvert}
{\lvert E_{\mathrm{expected}} \rvert}
\end{equation}

\begin{figure}[t]
\centering
\includegraphics[width=0.8\linewidth]{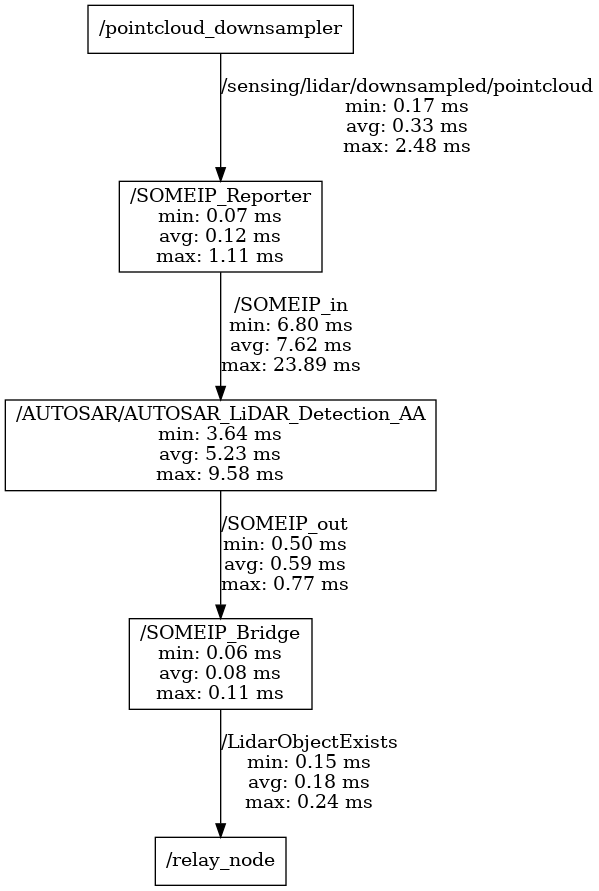}
\caption{Latency breakdown along the reconstructed end-to-end path (min/avg/max).}
\label{fig:chain_latency}
\vspace{-4mm}
\end{figure}

In this evaluation, the reconstructed topology achieved a coverage of 100\% with respect to the expected end-to-end graph.
This indicates that all expected dependency edges, including the cross-platform edges between ROS~2 and AUTOSAR~AP, were observed in the reconstructed graph.
Topology coverage is a diagnostic indicator that the cross-platform path was exercised, not a measure of behavioral correctness, scenario coverage, or safety.
This is visually corroborated by the end-to-end chain latency in Fig.~\ref{fig:chain_latency}: the reconstructed path extends without discontinuity from the ROS~2 sensor node~(\textit{/pointcloud\_downsampler}) to the actuation node, supporting the answer to RQ2.

\begin{figure}[t]
\centering
\includegraphics[width=\linewidth]{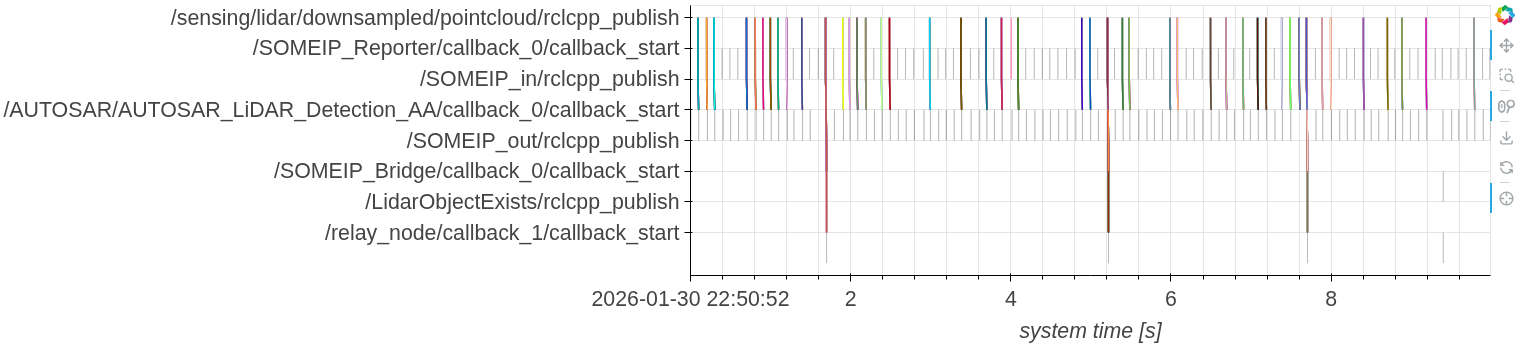}
\caption{End-to-end message flow reconstructed from merged ROS~2 and AUTOSAR~AP traces.}
\label{fig:message_flow}
\vspace{-5mm}
\end{figure}

The reconstructed message flow within the end-to-end path is visualized in Fig.~\ref{fig:message_flow}.
The reconstructed trace extends to the final actuation callback (\nolinkurl{relay\_node/callback_1/callback_start}) only when the object detection result is true (i.e., a stop command has been actively issued); when the result is false, the flow terminates earlier without reaching the actuation node.
This visual distinction shows that the reconstructed graph captures the dynamic scenario in which the obstacle disappears, the detection becomes false, and the ego-vehicle resumes motion, indicating that BA-TRACE traces varying control logic across the heterogeneous boundary in response to changes in the traffic situation.

The latency statistics in Fig.~\ref{fig:chain_latency} localize boundary-specific overhead along the reconstructed path.
The stages leading up to the Object Detection AA, specifically the \textit{/SOMEIP\_in} path, involve the transmission of heavy point cloud data, and the reconstructed graph attributes a relatively larger communication overhead~(avg: 7.62~ms) to this section compared with other sections.
The latency observed in \textit{/SOMEIP\_in} includes serialization, transport, and deserialization overhead, and should be interpreted as an aggregated communication cost across the DDS--SOME/IP boundary.
In contrast, the subsequent stages, which transmit only lightweight detection results or stop commands via \textit{/SOMEIP\_out}, exhibit minimal latency~(avg: 0.59~ms), an order of magnitude smaller than the /SOMEIP\_in path.
This ability to distinguish boundary-specific overhead according to data payload illustrates that BA-TRACE makes scenario-triggered behavior diagnostically interpretable in realistic mixed-platform environments.

These results support RQ3: the reconstructed graph turns external scenario outcomes into diagnostically interpretable timing evidence about internal cross-platform behavior at the DDS--SOME/IP boundary.
\section{Lessons Learned}
\label{chap:lessons_learned}

Based on the evaluation results, the three research questions of Section~\ref{sec:introduction} are revisited, followed by practical insights obtained through the design and operation of \emph{BA-TRACE}.
The discussion focuses on the role of boundary-aware trace reconstruction in interpreting scenario-triggered behavior of mixed AUTOSAR~AP--ROS~2 vehicular embedded systems.

\subsection{Validation of Research Questions}
\label{sec:lessons_learned_rq_validation}

Regarding \textbf{RQ1}, the evaluation made the traceability gap concrete.
Scenario-triggered data flow in the mixed stack crosses the DDS--SOME/IP boundary at the bridge converter, and ROS~2 trace events or AUTOSAR \texttt{ara::log} events alone cannot describe the cross-platform path.
ARXML-derived structural dependencies and bridge-level instrumentation were both required to fill this gap, confirming that the gap is structural rather than incidental to a particular scenario.

Regarding \textbf{RQ2}, BA-TRACE reconstructed an end-to-end execution graph aligned with the expected cross-platform processing topology, including the path through the point-cloud transfer stage, the AUTOSAR-side object-detection application, and the actuation-side return path.
This indicates that boundary-aware trace reconstruction can recover expected execution topology across DDS and SOME/IP domains.
The topology, however, is evidence of traceability rather than proof of behavioral correctness or safety.

Regarding \textbf{RQ3}, the reconstructed graph supported timing interpretation of scenario-triggered behavior. The latency breakdown localized boundary-specific overhead: point-cloud transfer dominated
the DDS-to-SOME/IP path, while the lighter detection-result return path was an order of magnitude smaller. In addition, the reconstructed message flow showed that the actuation path is reached only when the object-detection result is true and terminates earlier otherwise.
These results show that the reconstructed graph turns external scenario outcomes into diagnostically interpretable timing evidence about internal cross-platform behavior.

\subsection{Practical Insights}
\label{sec:lessons_learned_practical_insights}

\noindent\textbf{Bridge converters are observation points, not just protocol mappers:}
The bridge is the only component that simultaneously observes ROS~2-side topic semantics and AUTOSAR-side SOME/IP service semantics.
Treating it as an instrumentation point rather than a transparent translator is what makes cross-platform reconstruction tractable.

\noindent\textbf{Static and dynamic sources must be combined:}
Neither dynamic traces nor ARXML-derived structural dependencies are individually sufficient.
Dynamic events miss declared structure that is not exercised in a given scenario, while ARXML alone cannot describe runtime timing.
Combining them allowed expected dependency edges to be checked against observed ones.

\noindent\textbf{Topology coverage is a diagnostic indicator, not a safety metric:}
The fraction of expected dependency edges observed in the reconstructed graph indicates how completely a scenario exercised the cross-platform path.
It does not indicate scenario coverage, behavioral correctness, or safety, and should be reported with this scope.

\noindent\textbf{Single-instance deployment supports timing interpretability:}
Co-locating the simulator and the software stack on a single instance removes network-induced variability and lets boundary-specific latency be attributed to the software stack itself.
This is a methodological choice for diagnostic clarity, not a substitute for distributed-deployment evaluation.

\section{Related Work}
\label{chap:related_work}

\begin{table*}[t]
  \centering
  \caption{Comparison of validation methodologies}
  \label{tab:validation_comparison}
  \resizebox{\textwidth}{!}{%
  \renewcommand{\arraystretch}{0.95}
  \begin{tabular}{|l|c|c|c|c|c|}
    \hline
    Method / Paper & AUTOSAR~AP & ROS~2 & Validation Type & Scenario & Simulation \\
    \hline
    Real-world Testing~\cite{fei2024siltrack, fremont2020formal, allamaa2022nmpc}
      & \cm & \cm & Functional & Limited &  \\ \hline

    HIL Testing~\cite{abboush2024hilfi, vandeSluis2021hilv2x, chung2021hilmpc, devika2021hilcas, allamaa2022nmpc}
      & \cm & \cm & Real-time & Limited & Partially \\ \hline

    ASAM OpenSCENARIO~\cite{OpenSCENARIO} &  &  & Functional (Black-box) & \cm & \cm \\ \hline
    CARLA~\cite{CARLA} &  & \cm & Functional (Black-box) & \cm & \cm \\ \hline

    Logical scenario generation (ISO~34502-aligned)~\cite{jeon2023logical} &  &  & Methodology / Process & \cm & \cm \\ \hline
    Survey on safety-critical scenario generation~\cite{ding2022survey} &  &  & Survey / Taxonomy & \cm & Limited \\ \hline
    Scenario-based accelerated testing for SOTIF~\cite{tang2025sotifreview} &  &  & Survey / Taxonomy & \cm & Limited \\ \hline

    VIVAS~\cite{goyal2024vivas} &  & \cm & Functional / Formal & \cm & \cm \\ \hline
    SimValidation~\cite{li2023simvalidation} &  & \cm & Functional / Formal & \cm & \cm \\ \hline
    Formal scenario-based testing (sim-to-track)~\cite{fremont2020formal} &  & Limited & Formal & \cm & \cm \\ \hline

    SimADFuzz~\cite{xie2022simadfuzz} &  & \cm & Robustness (Black-box) & \cm & \cm \\ \hline
    Adaptive scenario generation~\cite{mullins2018adaptive} &  & \cm & Robustness (Search-based) & \cm & \cm \\ \hline
    Diversity-oriented exploration~\cite{ji2025diversity} &  & \cm & Robustness (Diversity-driven) & \cm & \cm \\ \hline
    SafeVar~\cite{pan2025safevar} &  & \cm & Robustness (Uncertainty) & \cm & \cm \\ \hline

    This paper & \cm & \cm & Performance / Interpretability (White-box) & \cm & \cm \\ \hline
  \end{tabular}}
  \vspace{-3mm}
\end{table*}

Prior work on scenario-based validation, autonomous driving simulators, and systematic testing provides the basis for practical scenario-driven validation of mixed AUTOSAR~AP--ROS~2 systems.
The comparison results are summarized in Table~\ref{tab:validation_comparison}.

\subsection{Scenario-based Validation Foundations}
\label{sec:related_work_foundations}

Scenario-based validation turns complex traffic situations into analyzable, reproducible artifacts and supports automated generation, execution, and analysis beyond mileage-oriented testing.
ISO~34502-aligned logical scenario generation defines parameter ranges for concrete tests~\cite{jeon2023logical}, while existing surveys and SOTIF-oriented studies organize generation, classification, risk-oriented testing, and accelerated assessment~\cite{ding2022survey,tang2025sotifreview}.
These studies ground scenarios as validation artifacts, although they emphasize design and generation rather than execution and internal interpretation in mixed software environments.

\subsection{Simulation-based Validation Frameworks}
\label{sec:related_work_frameworks}

Simulation-based validation frameworks automate scenario execution through simulators and standardized descriptions, with CARLA~\cite{CARLA} supporting controlled testing of perception, planning, and control stacks and ASAM OpenSCENARIO~\cite{OpenSCENARIO} enabling repeatable, interoperable dynamic scenarios.

VIVAS~\cite{goyal2024vivas}, SimValidation~\cite{li2023simvalidation}, and formal scenario-based testing~\cite{fremont2020formal} link scenario execution with coverage criteria, runtime monitoring, or safety properties.
These frameworks strengthen functional and formal assurance, although they evaluate observable outcomes and provide limited support for interpreting cross-platform internal behavior in heterogeneous software architectures.

\subsection{Search-based and Robustness-oriented Scenario Exploration}
\label{sec:related_work_search}

Falsification and search-based testing discover failure cases in scenario parameter spaces.
SimADFuzz~\cite{xie2022simadfuzz} uses simulation feedback and guided mutation to prioritize violation-inducing scenarios, while related methods characterize limits through adaptive search, diversity-oriented exploration, and uncertainty-aware analysis.
Adaptive search identifies performance boundaries and clusters~\cite{mullins2018adaptive}, diversity-oriented exploration promotes distinct failures~\cite{ji2025diversity}, and SafeVar shows physical-parameter variations, such as mass and friction, can increase unsafe outcomes~\cite{pan2025safevar}.

These studies support failure discovery and robustness assessment, although they differ from validation integrating scenario execution with white-box interpretation of mixed-platform behavior.

\subsection{Positioning of This Study}
\label{sec:related_work_positioning}

Existing scenario-based studies focus on generating, executing, or evaluating driving scenarios from an externally observable perspective, while existing tracing tools provide detailed observations within specific platforms such as ROS~2 or AUTOSAR~AP.
The gap addressed in this paper lies between these two directions: in mixed vehicular embedded systems, scenario-triggered behavior must be interpreted across heterogeneous middleware boundaries.
BA-TRACE bridges this gap by reconstructing scenario-triggered execution paths across DDS and SOME/IP, combining ROS~2 trace events, AUTOSAR \texttt{ara::log} events, ARXML-derived structural dependencies, and bridge-level instrumentation to provide diagnostic timing interpretability for mixed AUTOSAR~AP--ROS~2 vehicular embedded systems.

\section{Conclusion}
\label{chap:conclusion}

Modern vehicular embedded systems increasingly combine ROS~2-based autonomous-driving software with AUTOSAR~AP applications, but execution paths in such mixed stacks cross heterogeneous middleware boundaries that platform-local simulation and tracing tools cannot reconstruct on their own.
This paper presented \emph{BA-TRACE}, a boundary-aware trace reconstruction framework for scenario-based evaluation of mixed AUTOSAR~AP and ROS~2 vehicular embedded systems.
BA-TRACE combines ROS~2 trace events, AUTOSAR \texttt{ara::log} events, ARXML-derived structural dependencies, and bridge-level instrumentation to reconstruct an end-to-end execution graph across DDS and SOME/IP communication domains.

The AWSIM/OpenSCENARIO-based evaluation, conducted as an edge-case object-detection and braking case study, showed that BA-TRACE reconstructed the expected cross-platform processing path and related externally observed scenario behavior to internal timing behavior.
In particular, the reconstructed latency breakdown exposed boundary-specific communication overhead associated with point-cloud transfer across the DDS--SOME/IP boundary, while the lighter detection-result return path remained an order of magnitude smaller.
These results provided evidence of diagnostic traceability and timing interpretability for scenario-triggered behavior, not proof of behavioral correctness or safety.

Future work will extend BA-TRACE along three directions: broader scenario sets to test the generality of the reconstruction workflow, alternative component placements between AUTOSAR~AP and ROS~2 to assess sensitivity to deployment topology, and larger-scale regression testing environments.
A complementary direction is to combine BA-TRACE with automated scenario generation and search-based testing so that high-risk or low-coverage situations can be explored systematically across the DDS--SOME/IP boundary.

\section*{Acknowledgments}
This work was supported by JST FOREST Grant Number JPMJFR242G.

\bibliographystyle{unsrt}
\bibliography{refs}

\vfill

\end{document}